\documentclass[useAMS,letters,usenatbib]{mn2e}
\usepackage{epsfig,natbib}
\usepackage{myaasmacros,amssymb}

\newcommand{\dd}{\mathrm{d}}
\newcommand{\angst}{\mathrm{\AA}}
\def \mgii {{\rm Mg\,{\sc ii}~}}
\def \ciii {{\rm C\,{\sc iii]}~}}
\def \civ {{\rm C\,{\sc iv}~}}

\begin{document}

\title[Observational Test Rules Out Small \mgii Absorbers]{Direct Observational Test Rules Out Small \mgii Absorbers}
\author[A. Pontzen et al.]{Andrew Pontzen$^{1}$\thanks{Email: apontzen@ast.cam.ac.uk},
Paul Hewett$^{1}$,
Robert Carswell$^{1}$,
Vivienne Wild$^{2}$ \\
$^{1}$Institute of Astronomy, Madingley Road, Cambridge CB3 0HA, UK \\
$^{2}$Max-Planck-Institut f\"{u}r Astrophysik, 85748 Garching, Germany}
\date{Accepted 2007 August 5. Received 2007 June 26.}
\pubyear{2007}

\maketitle

\begin{abstract}
  Recent observations suggest the incidence of strong intervening
  \mgii absorption systems along the line-of-sight to gamma ray burst
  (GRB) afterglows is significantly higher than expected from
  analogous quasar sightlines. One possible explanation is a geometric
  effect, arising because \mgii absorbers only partially cover the
  quasar continuum regions, in which case \mgii absorbers must be
  considerably smaller than previous estimates.  We investigate the
  production of abnormal absorption profiles by partial coverage and
  conclude that the lack of any known anomalous profiles in observed
  systems, whilst constraining, cannot on its own rule out patchy
  \mgii absorbers.
  
  In a separate test, we look for differences in the distribution
  function of \mgii equivalent widths over quasar continuum regions
  and \ciii emission lines. We show that these anomalies should be
  observable in any scenario where \mgii absorbers are very small, but
  they are not present in the data. We conclude that models invoking
  small \mgii cloudlets to explain the excess of absorbers seen
  towards GRBs are ruled out.

\end{abstract}

\begin{keywords}
quasars: absorption lines
\end{keywords}

\section{Introduction}\label{sec:introduction}

The recent claim that strong \mgii absorption along the sightlines to
gamma ray bursts (GRBs) is significantly more likely than along the
sightlines to quasars \citep[][ hereafter P06]{2006ApJ...648L..93P} is
puzzling.  Subsequent results suggest that \mgii absorbers are unique
in exhibiting this trend, \civ absorbers displaying population
statistics independent of the nature of the background source
\citep{2007arXiv0705.0387T,2007arXiv0705.0706S}. One possibility is
that some of the \mgii absorbers seen towards GRBs are intrinsic,
although the high velocity separation between GRB and absorber coupled
with the low velocity dispersion in the observed \mgii components
prove a challenge to theoretical models. Therefore, it remains of
interest to investigate possible explanations for the effect in terms
of intervening absorbers.

This letter will focus on the model proposed by
\citet[][ hereafter F06]{2006astro.ph..5676F}, which suggests geometric
effects are responsible for the discrepancy. In this picture, the \mgii
absorbers are somewhat smaller than the beam size of the quasar continuum
region, whereas GRB beam sizes are smaller still.  This would
evidently lead to a difference in the observed distribution of
absorber properties.

However, such small absorber sizes do not fit well with the established
observational picture.  Information on absorber sizes comes from studies of
galaxies near quasar absorber sightlines, and from comparing absorber and galaxy
number densities.  These give typical radii $\sim$40\,kpc for \mgii complexes,
which may contain several velocity components, for a rest equivalent width limit
of 0.3\,\AA~\citep{1993ASSL..188..263S, 2007ApJ...658..161Z}.  For lower
equivalent width limits the inferred sizes are larger
\citep{1999ApJS..120...51C}, and projected \mgii absorber-galaxy correlation
functions show a significant excess out to over 1\,Mpc
\citep[e.g.][]{2004MNRAS.354L..25B}.  For individual velocity components of low
ionization species, sizes are less than $\sim$300--400\,pc
\citep{2002ApJ...576...45R}.

On the other hand, quasar continuum regions are generally considered
to be orders of magnitude smaller.  Important for the investigation
presented here is the significant difference in the physical size of
the ultraviolet continuum emitting region and the broad emission line
region in luminous quasars.  Constraints of the physical size of the
broad line and continuum regions are provided by reverberation studies
\citep[e.g.][]{2005ApJ...629...61K, 2007ApJ...659..997K} and
gravitational lensing studies \citep{1998MNRAS.295..573L}. Recently
the latter method was used by \citet{2005MNRAS.359..561W}, giving
values of $0.02$\,pc for the continuum region, with a broad-line
region $>3$ times larger (possibly considerably larger than this lower
limit).

A comprehensive analysis of possible routes to solving the GRB
discrepancy has recently been presented by \cite{2007astro.ph..1153P},
wherein it is suggested that the model in F06 has severe weaknesses.
These relate to the neglect of multiple clouds in the F06 model, and
the difficulty in producing doublet ratios which are consistent with
observations.  Whilst \cite{2007astro.ph..1153P} propose strong
objections to the model, they do not rule it out. Furthermore, the
idea appears to have been gaining popularity and has been used to
interpret recent observations \citep{2007ApJ...659L..99H}.

In this letter, we provide direct observational evidence against the
F06 model. First, we briefly discuss whether patchy absorbers can be
ruled out on the basis of their absorption profiles and line ratios
alone (Section \ref{sec:prof-single-comp}), and conclude that this is
not always the case.  We then use the differing size of the quasar
continuum and \ciii emission region to provide a definitive test for
the F06 model (Section \ref{sec:sdss-evid-against}). We provide a
summary in Section \ref{sec:conclusions}.

\section{Profiles of Single Velocity Component Partial
  Absorbers}\label{sec:prof-single-comp}

Suppose quasars emit light with a hard-edged circular beam, radius
$R_e$. Then a circular, single component \mgii absorber can be
parametrized by its central column density ($N_{\mathrm{MgII}}$),
velocity width ($v$) and its radius as a fraction of the beam radius
($R_0/R_e$).  In practice the `beam width' is an angular line-of-sight
effect and will scale accordingly as $R_0 = \theta D_A(z)$ where $z$
is the redshift of the absorber and $D_A$ is the angular diameter
distance. However, this geometric effect will be neglected as
$D_A(z_{\mathrm{abs}})\simeq D_A(z_{\mathrm{em}})$.  We set $R_e=1$
without loss of generality.

For $R_0<1$, providing that spectra of adequate S/N and velocity
resolution are obtained, saturated profiles with a systematic offset
above the zero-flux level (resulting from light `leaking' around the
edges) will be seen.  However, it is not realistic to model absorbers
with a hard edge; there must be a region around their edges where the
column density drops. In the absence of any favoured theoretical or
empirical model, we write:

\begin{equation}
N_{\mathrm{MgII}}(R)=\left\{\begin{array}{ll}
N_0 & R<R_0 \\
N_0\left(\frac{R}{R_0}\right)^{-n} & R>R_0
\end{array}\right. \mathrm{.}
\end{equation}

\noindent
During the drafting of this paper, a revised version of P06
independently proposed a similar model, using a three dimensional
distribution:

\begin{equation}
\rho(r)= \left\{\begin{array}{ll}
\rho_0 & r<r_0 \\
\rho_0\left(\frac{r}{r_0}\right)^{-k} & r>r_0
\end{array}\right. \mathrm{.}
\end{equation}

\noindent
This latter model is less convenient for our purposes of computing
high resolution profiles, something which P06 did not attempt.
However, one may roughly relate the two distributions simply by
integrating the second to yield $n=k-1$ outside the core radius.

The resulting observed profile is calculated numerically as:

\begin{equation}
F(\lambda) \propto \int_0^1 2\pi R e^{-\tau_\lambda(N_{\mathrm{MgII}}(R),v)}\dd R
\end{equation}

\noindent where $\tau_\lambda (N,v)$ is a standard Voigt optical 
depth profile for
\mgii$\lambda 2796$, and $F(\lambda)$ is the observed flux.  We
separately calculate the somewhat weaker profile \mgii$\lambda 2803$ so
that we can fit for both components simultaneously. We take $v=5$\,
kms$^{-1}$, which is the mean broadening for individual \mgii clouds
\citep{1997PhDT........18C}.

The model spectra are realized at high resolution ($\delta \lambda = 0.01
\angst$ in the absorber restframe). Gaussian random noise is added
with S/N=100, to approximate the best available echelle spectra. We
then fit a Voigt profile to the two components, assuming the correct
ratio of oscillator strengths. The resulting reduced $\chi^2$,
measured over the region where the departure from continuum level is
significant at the $1 \sigma$ level, gives an indication of how well
the partial absorber can fool one into believing the feature is due to
a full coverage absorber, including the effect of the doublet ratio.

We consider a grid of models of varying $N_{\mathrm{obs}}$ and $R_0$,
where $N_{\mathrm{obs}}$ is the observed (fitted) column density, not
the intrinsic column density of the partial absorber. This is achieved
by increasing, for each model, the intrinsic column density $N_0$
until the best fit gives the target $N_{\mathrm{obs}}$. The procedure
ensures each case we consider is a candidate for an observable system,
and automatically includes cases where $N_0$ is considerably larger
than is typically seen in \mgii absorbers but $R_0$ is very small,
which should be the most challenging case to disguise.

The results for $n=1,2$ (chosen for the reasons given below) are shown
in Figure 1 by dashed and solid lines respectively with $\log_{10}
N_{\mathrm{obs}}=12,14,16$ in red, green and blue. The shaded regions
indicate the 1$\sigma$ variance on the reduced $\chi^2$ for individual
fits: note that this depends on the number of pixels used in its
construction, and so will depend on $N_{\mathrm{obs}}$ (but not
significantly on $n$ or $R_0$, since $N_{\mathrm{obs}}$ is held
constant as these are varied).

\begin{figure}
\begin{center}
\includegraphics*[width=0.45\textwidth]{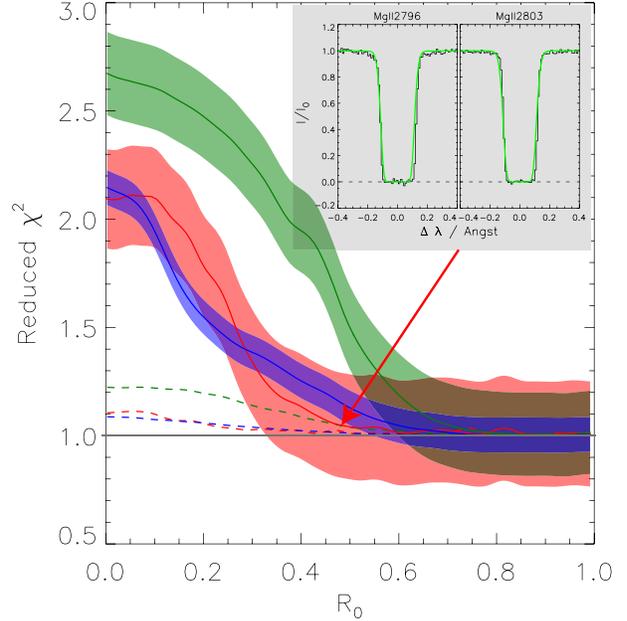}
\end{center}
\caption{The reduced $\chi^2$ of the best fit pure Voigt profile to a
  partial absorber with a core radius $R_0$ ($R_e=1$), outside of
  which the column density decays as $R^{-1}$ (dashed lines) and
  $R^{-2}$ (solid lines; the shaded regions indicate the variance of
  $\chi^2$ of individual fits, which is the same for the different $n$
  values but differs according to line strength because of the
  different number of pixels used in constraining the fit).  The red,
  green and blue lines indicate the results for {\it observed} column
  densities of $\log_{10} N_{\mathrm{obs}}=12, 14, 16$ respectively.
  Note this means the {\it intrinsic} central column density increases
  with decreasing radius.  As an example, the inset figure shows the
  profile for $R^{-2}$ law, $N_{\mathrm{obs}}=10^{14}, R_0/R_e=0.5$,
  with the best fit plotted in green.  By eye the partial coverage is
  virtually impossible to detect.}\label{fig:chi2-m5}
\end{figure}


F06 reproduce their geometric results for profiles as soft as
$n=1$ ($k=2$) with core radii $R_0 \sim 0.25$.  According to our
investigations, individual absorbers in this scenario would be
observationally indistinguishable from the complete coverage case,
with reduced $\chi^2 \sim 1$ for all fits down to $R_0 = 0.01$. As one
increases $n$ (making the absorber edges harder), it rapidly becomes
easier to distinguish the cases; for example, for $n=2$ and $R_0 \sim
0.25$, the high resolution spectra simulated here will highlight a
problem. Nevertheless, the $\chi^2$ values presented in Figure 1
represent the poorest possible quality of fit 
given that the model spectra are of exceptionally high
quality and only allow for fitting one velocity component.

Given the results described above, it may appear possible to have a
situation in which partial absorption is present but has not been
detected in the high resolution data. The physical explanation is
simply that the shape of the line depends on $N$ only logarithmically,
so allowing $N$ to vary as a shallow powerlaw across the beam does not
have a very strong effect: nonetheless, according to F06, one obtains
statistical variations in the equivalent width distribution according
to the size of the background quasar.

However, with such very slow powerlaw decays, there are further
problems to be investigated. On the one hand, the F06 statistical
result is valid only in a regime where the differing equivalent widths
of observed absorbers is generated by the differing radii of
intersection through a population of very similar clouds. On the other
hand, there is a very large area over which low column densities may
be observed on the periphery of such systems. In particular the
observed column density distribution of a single system may be
numerically calculated to scale as $\dd P(W)/ \dd W \sim W^{-3}$. It
seems unlikely that one could contrive a cosmological distribution of
absorbers which provide a rescue from such a steep powerlaw to give
the observed $\dd P(W) / \dd W \sim W^{-1}$ overall distribution
\citep[e.g.][]{2007ApJ...660.1093N,1999ApJS..120...51C} whilst
preserving the F06 result for $n=1$.

In conclusion, it may be that soft edges can be invoked to alleviate
somewhat the high resolution fitting problems associated with patchy
absorption. However, whilst the limit of extremely shallow outer
slopes yields very convincing Voigt profiles, one obtains a new set
of statistical challenges. For the rest of this letter, we will assume
sharper cut-offs which do not suffer from these problems.

\section{\mgii absorbers close to
\ciii$\lambda 1909$ emission in quasar spectra}\label{sec:sdss-evid-against}

If, as in F06, \mgii systems are somewhat smaller than quasar
continuum beam widths, it follows that they cannot cover the broad
emission line region (Section \ref{sec:introduction}), except with a
much reduced column density which must be negligible according to the
statistical argument at the end of Section
\ref{sec:prof-single-comp}. Accordingly, one can approximately model
the spectrum over a broad emission line as:

\begin{equation}
F(\lambda)=C(\lambda)e^{-\tau_{\mathrm{MgII}}(\lambda)}+L(\lambda)\label{eq:spec-mod}
\end{equation}

\noindent where $F$ is the received flux, $C$ is the continuum
emission flux, $L$ is the broad-line emission flux and
$\tau_{\mathrm{MgII}}$ is the optical depth due to an intervening
\mgii system.
Assuming~(\ref{eq:spec-mod}), an absorber with ``intrinsic'' equivalent
width $W$ (as seen over the continuum region only) will be seen with
equivalent width $W'$:

\begin{equation}
W'=\frac{W}{1+L/C}\label{eq:1}
\end{equation}

\noindent where $L(\lambda)$ and $C(\lambda)$ are assumed to be
constant over the absorption feature (compare typical widths of
20\,kms$^{-1}$ and 5000\,kms$^{-1}$ for a \mgii component and quasar
broad line emission respectively). This result applies for any
absorber size that is smaller than the continuum region. Possible
complications are discussed in Sections \ref{sec:compl-a:-centre} \&
\ref{sec:compl-b:-mult}, but essentially this simple model is adequate
for our investigation.  Equation (\ref{eq:1}) implies that, for small
absorbers, the observed number of absorbers above a specified
equivalent width threshold may be reduced significantly over broad
line emission features because of the `diluting' effect of the light
from the large broad emission line region.  It is thus possible to
apply a test similar to the geometric intrinsic/extrinsic test of the
Ly$\alpha$ forest described in \cite{1980ApJS...42...41S}.

\subsection{Observational Sample}

The \mgii\,$\lambda\lambda$2796,2803 absorber sample of \citet[][
Section 2]{2007MNRAS.374..292W} is used as the basis for the
statistical investigation of the observed absorber properties in the
vicinity of the \ciii\,$\lambda 1909$ emission line.  It is derived
from a S/N-limited search of 32\,278 quasar spectra contained in the
Sloan Digital Sky Survey (SDSS) Data Release 4 (DR4)
\citep{2006ApJS..162...38A}.  The sample includes absorbers down to a
restframe equivalent width ($W$) of 0.5\,\AA \ over an extended
redshift range.  For our purposes we confine attention to the 2\,831
absorbers with $W_{0}^{2796}\ge$0.9\,\AA~ and $-180<\delta
\lambda/$\AA~$<180$ where $\delta \lambda$ is the wavelength shift
between the emission line and absorption line in the quasar restframe:

\begin{equation}
\delta \lambda = \frac{2796\angst(1+z_{\mathrm{MgII}})}{(1+z_{\mathrm{quasar}})} - 1909\angst\textrm{ .}
\end{equation}

\noindent
Here $z_{\mathrm{MgII}}$ is the redshift of the absorber and $z_{\rm
  quasar}$ is the redshift of the quasar. The $\pm180$\,\AA \
wavelength interval was chosen to provide sufficient absorption lines
in the comparison sample. The $W_{0}^{2796}\ge$0.9\,\AA \ limit was
chosen to provide an absorber sample consisting of a large number of
strong \mgii absorbers but our conclusions are insensitive both to the
precise $W_{0}$-limit employed or the wavelength extent used to define
the comparison sample.

For the test below, we need to predict the absorbers as a function of
$\delta \lambda$, without the correction described by equation
(\ref{eq:1}) which will be considered separately. The expected number
will be a function of (a) the accessible redshift path, (b) the mean
redshift of the absorbers and (c) the S/N in the quasar spectra.  The
accessible redshift path as a function of $\delta\lambda$ is readily
calculated using the redshifts of the base sample of 32\,278 quasars.
The small systematic change in the number of absorbers due to
evolution with redshift can be calculated from the mean absorber
redshift as a function of $\delta\lambda$, combined with the absorber
number density as a function of redshift, $\propto(1+z_{\rm
  MgII})^{0.23}$, determined by \citet{2005ApJ...628..637N}. The
probability of detecting an absorption line of fixed equivalent width
varies as a function of S/N. The `signal' depends on the systematic
change in the height of the quasar continuum+\ciii\,$\lambda 1909$
emission line with $\delta\lambda$.  The `noise' varies due to a
combination of the increasing sky-background as a function of
observed-frame wavelength and of the sensitivity of the SDSS
spectrograph and detector. By combining these considerations we can
predict the number density of absorbers as a function of $\delta
\lambda$.


\subsection{Statistical Test}\label{sec:statistical-test}

\begin{figure}
\begin{center}
\includegraphics*[width=0.49\textwidth]{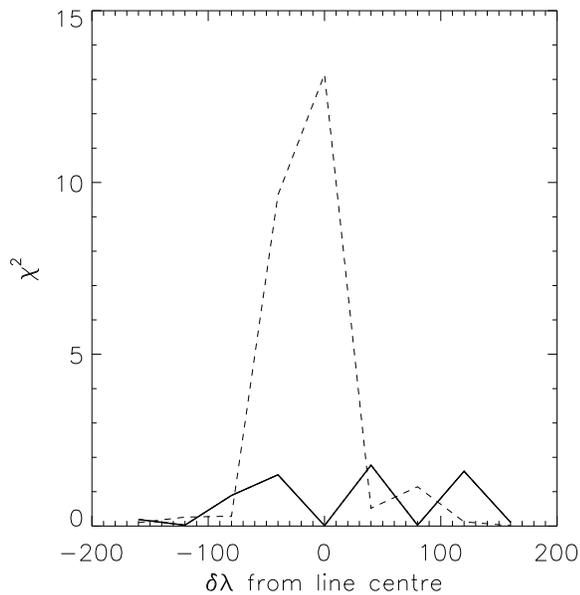}
\end{center}
\caption{The results of testing for a significant difference between
  the observed number of absorbers relative to the location of the
  \ciii\,$\lambda 1909$ emission line.  The thick solid line shows
  $\chi^2$ for the pure data. The dotted solid line shows $\chi^2$
  once the observations are corrected for the partial coverage effect
  (equation \ref{eq:1}); this results in more absorbers entering our
  sample, changing $\chi^2$. The original data is consistent with
  predictions, unlike the corrected data, so that we conclude the
  correction is fictitious.}\vspace{-0.3cm}\label{fig:chi2}
\end{figure}

We adopt a very simple statistical test, comparing the predicted
number of absorbers with $W_0^{2796}\ge$0.9\,\AA~
against the observed number, yielding a value of
$\chi^2=(N_{\mathrm{bin}}-N_{\mathrm{exp}}(\delta
\lambda))^2/N_{\mathrm{exp}}(\delta \lambda)$ in bins of 40\,\AA,
where $N_{\mathrm{bin}}$ is the observed number of absorbers in the
bin and $N_{\mathrm{exp}}(\delta \lambda)$ is the expected number of
absorbers in the bin according to the procedure detailed above. 
The bin size is chosen to maximize the number of absorbers per bin
whilst ensuring the effect of the broad emission line is not diluted by
choosing the bin width too large. The central bin is centred on
$\delta \lambda=0$.

Since one expects, in the absence of partial coverage,
$N_{\mathrm{bin}}$ to be Poisson distributed with mean
$N_{\mathrm{exp}}$, and $N_{\mathrm{bin}} \sim 300 \gg 1$, the
distribution should be accurately Gaussian and $\chi^2$ should be
$\sim 1$ in all bins. However, in the small absorber scenario, the
dilution effect will cause a drop in the number of observed absorbers
in the vicinity of the broad emission line, and hence an increase in
$\chi^2$.

The result is shown by the thick, solid line in Figure \ref{fig:chi2}.
The mean $\chi^2 \simeq 0.7$ indicates a result consistent with the
full coverage picture.  To investigate whether this could occur simply
because the effects we look for are too small, we now perform a second
test, in which we make the assumption that the observed equivalent
widths are altered over the \ciii emission line due to partial
coverage. In this picture, we have observationally measured $W'$, so
each absorber is then adjusted according to (\ref{eq:1}) to yield
$W$. $C$ is determined by interpolating continuum points from each
side of the \ciii feature, whilst $L+C$ and hence $L$ may be estimated
by inspecting the signal to either side of the \mgii feature. This
results in a number of absorbers, which previously had equivalent
widths below the 0.9\,\AA ~limit, entering our sample. The correction
factor to equivalent widths of absorbers on top of the emission
feature peaks at $1+L/C \simeq 1.5$, although there are a few outliers
where this factor takes a value as high as $1+L/C \simeq 2.1$.

Performing the statistical test yields an exceptionally poor fit
between the observations and the model predictions, shown by the
dashed line in Figure \ref{fig:chi2}.
We conclude that the partial coverage scenario is inconsistent with
the observations at a very high level of significance, certainly in
the simple limit considered here. We now show that the comparison
$\chi^2$ results are, in fact, not invalidated by our
oversimplification of the absorber model.



\subsection{Complication A: Off-Centre Single Absorber}\label{sec:compl-a:-centre}

Suppose the continuum region has area $A_c$, the absorber $A_a \le
A_c$ and the broad emission line region has area $A_l$.  If the area 
overlap between
the absorber and continuum is $A_0$, with total absorber area $A_a$,
it is simple to show that the observed equivalent width $W'$ is given
in terms of the intrinsic equivalent width $W$ by:

\begin{equation}
W'=\frac{A_0A_lC + (A_a-A_0)A_cL}{A_cA_l(C+L)}W\label{eq:2}\textrm{ .}
\end{equation}
  
Since one could never observe the intrinsic $W$ in this picture, the
correct quantity to consider is $W'(L,A_0)/W'(L=0,A_0)$, i.e. the
ratio between the equivalent width observed over an emission line to
that observed where there is no emission from the line region:

\begin{equation}
\frac{W'(L,A_0)}{W'(L=0,A_0)} = \frac{A_0A_lC + (A_a-A_0)A_c L}{(C+L)A_0A_l}\textrm{ .}
\end{equation}

For $A_0=A_a$, so that $W'(L=0,A_0)=W$, this equation correctly
reduces to equation (\ref{eq:1}).  However its behaviour elsewhere is
somewhat subtle; in the limit $A_0 \to 0$, one obtains
$W'(L,A_0)/W'(L=0,A_0)\to\infty$, apparently compensating for our
original effect and generating extra high $W$ absorbers.  However,
inserting the advocated $A_c \gtrsim 4 A_a$, and the observational
values $A_l\gtrsim 10 A_c$, $L \sim 0.2C$, one can use (\ref{eq:2}) to
show that the population of absorbers generating this compensating
effect must have intrinsic $W \gtrsim 250 W'$. Thus those absorbers
ever to enter our sample ($W'>0.5$\AA)~would have physically absurd
high intrinsic column densities. We conclude the effects are only
significant for $W'\ll 0.5$\,\AA, so that our statistical result
(Section \ref{sec:statistical-test}) is insensitive to these
considerations.


\subsection{Complication B: Multiple Absorbers}\label{sec:compl-b:-mult}

It is clear from the multi-component nature of observed \mgii systems
that any single cloud approach is over-simplified. If one asserts the
{\it overall} size of the multi-cloud complex is smaller than the
continuum region, our result holds identically since we have simply
shown that there is no detectable net difference in the fraction of
light absorbed against the continuum and \ciii line regions.

A separate scenario is multiple clouds per sightline, {\it each} with
$A_a \sim A_c$. This would introduce absorption over the broad
emission line, thus reducing the magnitude of our comparison
$\chi^2$. In \citet{2007astro.ph..1153P}, it is stated that the number
of clouds will scale with the beam area. This may or may not be a good
approximation to the true behaviour; however, in our opinion the
effect cannot be addressed in detail without a sophisticated model
which includes considerations of not only the abundance but
necessarily the spatial correlation of \mgii systems.

For instance, in a model where \mgii systems cluster in spheres of a
characteristic radius, the total number of absorbers over the line
region will scale as $r^2/A_c$ where $r$ is the radius of the cluster
of absorbers, for $A_L > \pi r^2 > A_c$, but as $A_L/A_c$ for $\pi
r^2>A_L$. This latter scaling is also applicable in the case that
\mgii systems are considered to be evenly scattered across the local
patch of sky.

Discussion of these models are beyond the scope of this letter. But
such considerations do not affect our conclusions with regard to the
small \mgii cloud models. The models of F06 are calculated considering
only isolated clouds, and already these clouds must be adjusted in
size to fit the data. In order to simultaneously fit our negative
result and the GRB/quasar mismatch, one would need additionally to
model the distribution of clouds so that the approximation breaks down
in the correct manner and at the correct scale to compensate precisely
for the geometric dilution in our specific case. Such a degree of fine
tuning would not be acceptable without the support of a physically
motivated model.

\section{Conclusions}\label{sec:conclusions}

We investigated the possibility that \mgii absorber clouds are somewhat
smaller than, but comparable to, the beam width of the quasar continuum.
Such a scenario is motivated by disparate statistics of \mgii
absorption when comparing GRB and quasar sightlines
(F06).

All our results suggest that, without extreme fine-tuning, the model
is hard to match with the known properties of \mgii systems. High
resolution spectroscopy directly rules out partial coverage clouds
with hard edges, whilst softening the edge will mimic a pure Voigt
profile well but introduce difficulties with producing the known
equivalent width distribution. Most convincingly, a statistical test
for systematically lowered equivalent widths over quasar broad line
emission regions (which are substantially larger than quasar continuum
regions) produces no result.  After applying a correction on the
assumption that partial coverage is the correct scenario, an
inconsistency between the equivalent width distributions over broad
emission line and continuum regions arises.  The only circumstances in
which the F06 model holds and our broad emission line effect might be
unobservable are those in which extreme fine tuning of the sizes and
distributions are invoked (Section \ref{sec:compl-b:-mult}).

We therefore reject the proposition of \mgii absorbers which are
patchy on scales comparable to quasar continuum beam sizes. The true
explanation for the observations of \cite{2006ApJ...648L..93P} remains
elusive.

\section*{Acknowledgments}
We thank Max Pettini and Berkeley Zych for useful comments on a draft
of this letter.  AP is supported by a STFC (formerly PPARC)
studentship and scholarship at St John's College, Cambridge. VW is
supported by the MAGPOP Marie Curie EU Research and Training Network.

\bibliographystyle{mn2e} \bibliography{/home/app26/documents/refs}

\end{document}